\UseRawInputEncoding
\documentclass[superscriptaddress, notitlepage, reprint,twocolumn]{revtex4-1}
\usepackage[english]{babel}
\usepackage{amsmath,amsthm,amssymb,lipsum}
\usepackage{amsfonts}
\usepackage[pdfborder={0 0 0}, colorlinks=true, urlcolor=blue, linkcolor=blue, citecolor=blue]{hyperref}
\usepackage{color}
\usepackage{graphicx}
\usepackage{float}
\usepackage{subfigure}
\usepackage{lipsum}
\usepackage{epstopdf}
\usepackage{dcolumn}
\usepackage{mathrsfs}
\begin{document}

\title{Gravimetry enhanced by nonreciprocal optomechanical coupling }
\author{Dong  Xie}
\email{xiedong@mail.ustc.edu.cn}
\author{Chunling Xu}

\affiliation{College of Science, Guilin University of Aerospace Technology, Guilin, Guangxi 541004, People's Republic of China}

\begin{abstract}
We explore how to measure the gravitational acceleration by using a dissipative optomechanical cavity. What is quite different from the conventional measurement methods is that we have constructed a nonreciprocal optomechanical coupling. We demonstrate that nonreciprocal coupling has a much greater advantage than reciprocal coupling. When the coherent optomechanical coupling is relatively weak and the driving intensity of single photon is strong, nonreciprocal coupling can improve the measurement precision by a factor of two. When the driving strength of single photon tends to infinity, reciprocal coupling fails to obtain any information about the gravitational acceleration, while nonreciprocal coupling still does. Using a two-photon driving, the measurement uncertainty of the gravitational acceleration will tend to zero as the intensity of two-photon driving approaches the critical point. The critical value of the two-photon driving intensity required for nonreciprocal coupling is finite, but the critical value of the two-photon driving intensity required for reciprocal coupling is infinite.  The combination of the amplification of mechanical parameters and the extra force can not improve the measurement precision, but it can enhance the susceptibility. Furthermore, when the single-photon driving field is relatively weak, we analytically calculate the quantum Fisher information. The results show that, under most experimental parameters, nonreciprocal coupling still performs better than reciprocal coupling.
\end{abstract}
\maketitle

\textit{Introduction}-The development of a highly precise gravimeter plays a very important role in both basic science and industrial applications.
Precise measurement of the gravitational acceleration can verify the accuracy of Newton's theory of gravitation and thereby promote the in-depth development of the quantum gravity theory\cite{lab1}. Gravimeter can assist in conducting tests of the principle of equivalence\cite{lab2}, and developing the theory of relativistic quantum metrology\cite{lab3,lab4,lab5,lab6}.
In terms of industrial applications,
precision gravimeters have been used in inertial navigation
technologies and for conducting geological surveys\cite{lab7,lab8,lab9,lab10}.

The centers-of-mass of massive mechanical objects are inherently subjected to gravitational acceleration. The information of gravitational acceleration $g$ can be naturally encoded into the mechanical modes. Therefore, massive mechanical object is an important component of most gravimeters\cite{lab11}. Massive mechanical object can be straightforwardly coupled to light-matter systems. Then, the information about gravitational acceleration is transferred to light-matter systems that are easier to measure. Furthermore, quantum effects of light-matter systems can be utilized to enhance the measurement precision, such as, a Heisenberg-limited spin-mechanical gravimeter\cite{lab12}.

Thanks to the very particular light-matter interaction, optomechanical cavity has become a promising platform for developing quantum sensors of the gravitational acceleration\cite{lab13,lab14,lab15,lab16,lab17,lab18,lab19,lab20,lab21}. Cavity optomechanics can be achieved through different methods in experiments, such as, a moving-end mirror\cite{lab22}, Brillouin scattering\cite{lab23}, nanomechanical rotors\cite{lab24}, whispering-gallery-mode optomechanics
\cite{lab25}, superconducting devices\cite{lab26}, and levitated sphere\cite{lab27}.
Recently, nonreciprocal coupling has been used to improve the sensing sensitivity of a driving
signal\cite{lab28}. This prompts us to further investigate whether nonreciprocal coupling can enhance the measurement precision of the gravitational acceleration. This is an area that has not been explored so far and it will provide new resources for further improving the measurement precision of the gravitational acceleration.

In this letter, we utilize an adiabatical elimination to construct a nonreciprocal coupling between the cavity-field mode and the mechanical mode. By measuring the steady-state cavity field, information about the gravitational acceleration can be obtained.
When the driving intensity of single-photon is strong, the mean-field approximation can be used to obtain analytical results about the measurement precision of the gravitational acceleration. When the coherent optomechanical coupling strength is relatively weak, we prove that nonreciprocal coupling can achieve a two-fold improvement in measurement precision compared to reciprocal coupling.
When the driving intensity of single-photon tends to infinity, the nonreciprocal coupling can obtain the measurement information of the gravitational acceleration, while the reciprocal coupling can not obtain any information.
By using two-photon driving, the optomechanical system will change from the steady state to the non-steady state as the driving intensity of two-photon increases. The phase transition from the steady state to the non-steady state appears at the critical point.  The measurement uncertainty of the gravitational acceleration will tend to zero as the intensity of two-photon driving approaches the critical point.
For nonreciprocal coupling, the critical two-photon driving intensity is a finite value; while for reciprocal coupling, the critical two-photon driving intensity is infinitely large. Therefore, nonreciprocal coupling can effectively utilize the critical point to reduce the measurement uncertainty of the gravitational acceleration to 0, while reciprocal coupling can not achieve this.
The combination of the amplification of mechanical parameters and the extra force can improve the susceptibility, i.e., the strength of the signal. However, at the same time, it also brings about the same noise intensity, thus making it impossible to directly enhance the signal-to-noise ratio. In addition, we discuss about the advantage of the nonreciprocal coupling in the case of
weak single-photon driving field. By truncating the quantum system in a low-dimensional space, we calculate the quantum Fisher information. As a result, we show that non-reciprocal coupling still
performs better than reciprocal coupling under most experimental parameters.

\begin{figure}[h]
\includegraphics[scale=0.32]{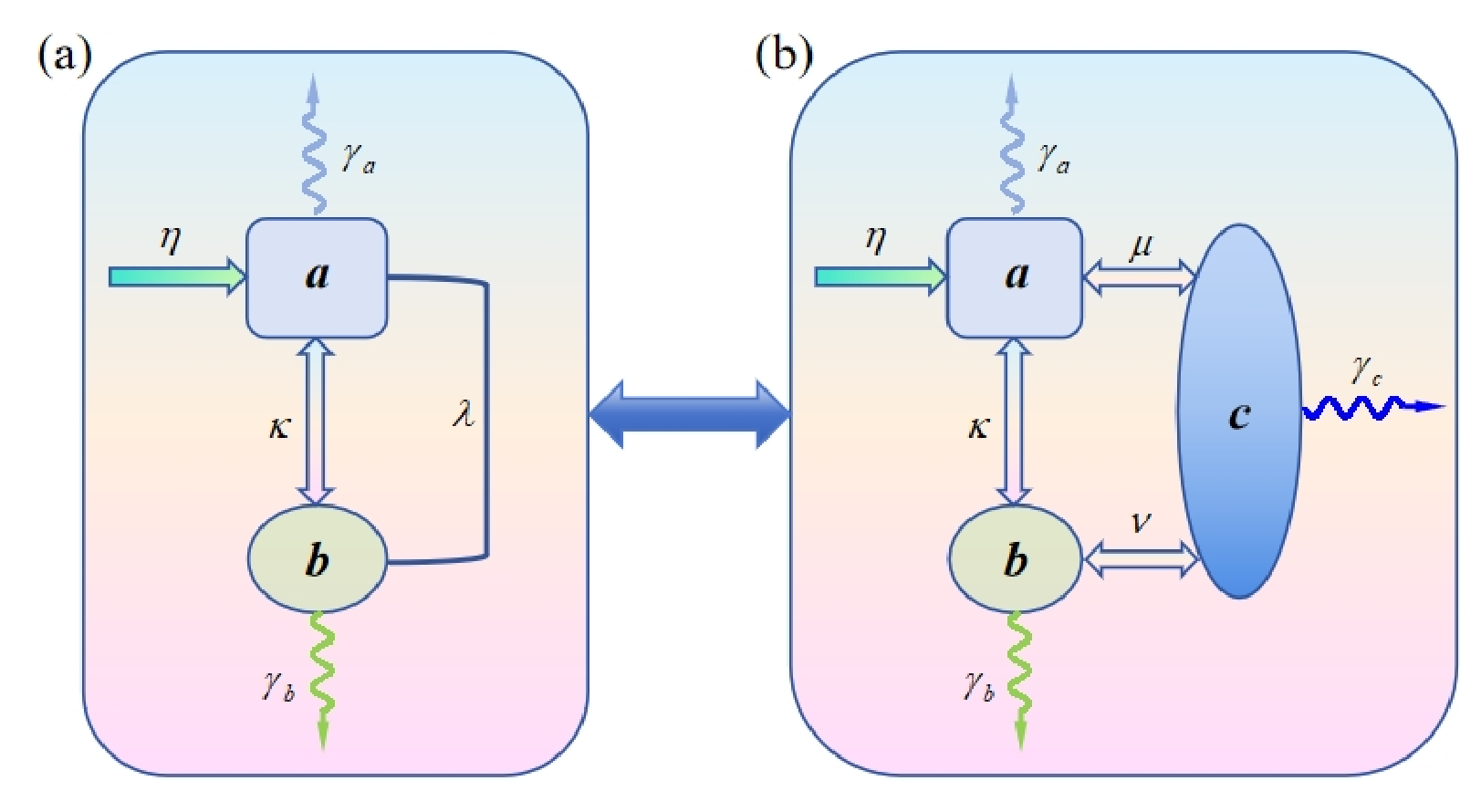}
 \caption{\label{fig.1} (a) A dissipative optomechanical system consists of a cavity-field mode $a$ and a
mechanical mode $b$. The cavity mode interacts with the mechanical mode by the coherent optomechanical coupling with the strength $\kappa$ and the dissipative coupling with the strength $\lambda$. They inevitably suffer local dissipative processes with the respective dissipation rates of $\gamma_a$ and $\gamma_b$. (b) The dissipative coupling can be achieved by coupling the optical-mechanical system to a dissipative mechanical mode $c$ with the dissipation rate $\gamma_c$. Here, $\eta$ denotes the single-photon driving strength.}
\end{figure}

 \textit{Optomechanical model}-We consider a general optomechanical system consisted of a mechanical oscillator coupled to a light-field in the cavity, as shown in Fig.~\ref{fig.1}.
The optomechanical Hamiltonian is described by ($\hbar=1$)\cite{lab16}
 \begin{align}
H&=\nonumber\\
&\omega_c a^\dagger a+\omega_b b^\dagger b-\kappa a^\dagger a(b^\dagger+ b)+\cos(\theta) g \sqrt{\frac{m}{2\omega_b}}(b^\dagger+ b),
\end{align}
where $a$ ($a^\dagger$) is the annihilation (creation) operator for the cavity field with frequency $\omega_c$, and $b$ ($b^\dagger$) is the annihilation (creation) operator for the mechanical oscillator with frequency $\omega_b$ and the $m$ is the mass of the
mirror, $g$ is the local gravitational acceleration and $\kappa$ is a coherent coupling constant that determines the interaction strength between the photon number and
the position of the oscillator. $\theta$ denotes the angle from the vertical axis and is used to describe the degree of inclination of the system. In order to achieve a high measurement precision of the gravitational acceleration $g$, $\theta$ is set to 0 throughout the following discussion.

The cavity-field mode and the mechanical mode suffer from the inevitable dissipation, which can be described by
the Lindblad master equation
\begin{align}
\dot{\rho}=-i[H,\rho]+\sum_{j=a,b}\gamma_j\mathcal{D}_j[\rho]+\lambda\mathcal{D}_z[\rho],\tag{2}\label{eq:2}
\end{align}
where the dissipation superoperator is $\mathcal{D}_o[\rho]=o\rho o^\dagger-\frac{1}{2}\{o^\dagger o, \rho\}$, $\gamma_j$ denotes the local dissipation rate,
and $\lambda$ denotes the dissipative coupling. We assume that the cavity system and the mechanical system are coupled to the common  reservoir with the same strength, i.e., the collective annihilation operator $z=(ia^\dagger a+b)$.
The dissipative coupling can be obtained through the interaction between the optomechanical system and a common mechanical harmonic oscillator, as shown in Fig.1(b). By choosing appropriate coupling strengths $\mu$ and $\nu$, as well as dissipation rate $\gamma_c$, and eliminating the dissipation mode $c$ through an adiabatic process, the dissipative coupling can be equivalently obtained (see the detail in Appendix: A ).

The Lindblad master equation can be mapped onto the corresponding quantum Langevin-Heisenberg equation of an operator $O$, which is derived by\cite{lab30,lab31,lab32}
\begin{align}
\dot{O}=&i[H,O]-\sum_{d=a,b,z}\{[O,d^\dagger](\kappa_dd-\sqrt{2\kappa_d}d_{{in}})\nonumber\\
&-(\kappa_dd^\dagger-\sqrt{2\kappa_d}d^\dagger_{{in}})[O,d]\},
\label{eq:3}
\tag{3}
\end{align}
where the expected values of noise operators $d_{{in}}=\{a_{{in}},\ b_{{in}}, z_{{in}}\}$ satisfy the relations
\begin{align}
&\langle d_{{in}}(t) \rangle=\langle d_{{in}}^\dagger(t) \rangle=0,\
\langle d^\dagger_{{in}}(t)d_{{in}} (t')\rangle=0,\label{eq:4}
\tag{4}\\
&\langle d_{{in}}(t)d^\dagger_{{in}} (t')\rangle=\delta(t-t').
\label{eq:5}
\tag{5}
\end{align}

In order to get a nonzero steady state, we use a single-photon driving field, which is described by
\begin{align}
H_d=\eta(ae^{i\omega_dt}+a^\dagger e^{-i\omega_dt}).
\label{eq:6}
\tag{6}
\end{align}
In order to obtain a large number of photons in the optical cavity, we have always considered that the driving frequency should satisfy the resonance condition: $\omega_d=\omega_c$.

In the frame rotating at $\omega_c$, the corresponding quantum Langevin-Heisenberg equations of operators $a$ and $b$ are given by
\begin{align}
\dot{a}=&(-\gamma_a -\lambda)a+i(\kappa+\lambda)a(b+b^\dagger)-i\eta \nonumber\\
&+\sqrt{2\gamma_a} a_{in}-i\sqrt{2\lambda}a(z_{in}+z^\dagger_{in}),\label{eq:7}
\tag{7}\\
\dot{b}=&(-i\omega_b-\gamma_b-\lambda )b+i(\kappa-\lambda) a^\dagger a -i G\nonumber\\
&+\sqrt{2\gamma_b} b_{in}+\sqrt{2\lambda}z_{in},
\label{eq:8}
\tag{8}
\end{align}
where the parameter $G=g\sqrt{m/2\omega_b}$.
When $\lambda=\kappa$, the evolution of the mode $b$  is not influenced by the mode $a$. In this case, we define the effective coupling between system $a$ and system $b$ to be completely non-reciprocal. When there is no dissipative coupling, i.e., $\lambda=0$, the interation between system $a$ and system $b$ is reciprocal.

\textit{Nonreciprocal coupling}:
In the case of a strong single-photon driving field, the nonlinear set of equations displayed in Eqs.(\ref{eq:7}-\ref{eq:8}) can be linearized by splitting the operators in terms of their mean values and quantum fluctuations as: $a=\langle a\rangle+\delta a$, and $b=\langle b\rangle+\delta b$. Then the evolution equation of the mean values  can be rewritten as
\begin{align}
\dot{\langle a\rangle}&=(-\gamma_a -\kappa)\langle a\rangle+2i\kappa \langle a \rangle\langle b+b^\dagger\rangle-i\eta,\label{eq:9}
\tag{9}\\
\dot{\langle b\rangle}&=(-i\omega_b-\gamma_b-\kappa )\langle b\rangle -i G.
\label{eq:10}
\tag{10}
\end{align}
Let the left terms $\dot{\langle a\rangle}=\dot{\langle b\rangle}=0$, the mean values at the steady state are derived
\begin{align}
\alpha=&{\langle a\rangle}=\frac{-i\eta}{\gamma_a +\kappa+4iG\kappa \omega_b/[(\gamma_b+\kappa)^2+\omega_b^2]},\label{eq:11}
\tag{11}\\
\beta=&{\langle b\rangle}=-i G/(i\omega_b+\gamma_b+\kappa ).
\label{eq:12}
\tag{12}
\end{align}

And the evolution equations of quantum fluctuations is given by
\begin{align}
\dot{\delta a}=&[-\gamma_a -\kappa+2i\kappa (\beta+\beta^*)]\delta a-2i\kappa \alpha( \delta b+\delta b^\dagger)\nonumber\\
 &+\sqrt{2\gamma_a} a_{in}-i\sqrt{2\lambda}\alpha(z_{in}+z^\dagger_{in}),\label{eq:13}
\tag{13}\\
\dot{\delta b}=&(-i\omega_b-\gamma_b-\kappa )\delta b +\sqrt{2\gamma_b} b_{in}+\sqrt{2\kappa}z_{in}.
\label{eq:14}
\tag{14}
\end{align}

The measurement uncertainty of the local gravitational acceleration $g$ is derived by the error propagation formula
\begin{align}
\delta g=\frac{\sqrt{\langle M^2\rangle-|\langle M\rangle|^2}}{|\partial_g\langle M\rangle|},
\label{eq:15}
\tag{15}
\end{align}
where $|\partial_g\langle M\rangle|$ denotes the susceptibility of the measurement signal to the gravitational acceleration.
When the single-photon driving field is strong and the steady state is Gaussian, the homodyne detection with the measurement operator $M=a+a^\dagger$ is close to the optimal measurement and easy to implement experimentally\cite{lab33,lab34,lab35}.
Then, we can obtain the signal and the noise from the homodyne detection
\begin{align}
&|\partial_g\langle M\rangle|=\frac{\eta G_1}{g((\gamma_a +\kappa)^2+G_1^2)}-\frac{2\eta G^3_1}{g((\gamma_a +\kappa)^2+G_1^2)^2},\label{eq:16}
\tag{16}\\
&\langle M^2\rangle-|\langle M\rangle|^2=1+2\langle\delta a^\dagger\delta a\rangle_s,\label{eq:17}
\tag{17}
\end{align}
where $G_1=4G\kappa \omega_b/[(\gamma_b+\kappa)^2+\omega_b^2]$ and $\langle\delta a^\dagger\delta a\rangle_s$ denotes the expected value at the steady state. Substituting the above equations into Eq.~(\ref{eq:15}), the analytical form of the measurement uncertainty can be obtained(see the detail in Appendix: B).

\textit{Reciprocal coupling}-When there is no dissipative coupling, i.e. $\lambda=0$, the quantum Langevin-Heisenberg equations are given by
\begin{align}
\dot{a}&=-\gamma_a a+i\kappa a(b+b^\dagger) +\sqrt{2\gamma_a} a_{in}-i\eta,\label{eq:18}
\tag{18}\\
\dot{b}&=(-i\omega_b-\gamma_b)b+i\kappa a^\dagger a +\sqrt{2\gamma_b} b_{in}-i G.
\label{eq:19}
\tag{19}
\end{align}
When $\kappa |\alpha|^2\ll G$, the measurement uncertainty of the gravitational acceleration can be derived. As a result, the ratio of the measurement uncertainty in both reciprocal and nonreciprocal cases can be obtained
\begin{align}
R=\frac{\delta g_{nr}}{\delta g_r}=\frac{1}{2},
\label{eq:20}
\tag{20}
\end{align}
where $\delta g_{nr}$ ($\delta g_r$) denotes the measurement uncertainty of the gravitational acceleration in the case of nonreciprocal (reciprocal) case.
According to the above equation, the measurement precision of nonreciprocal coupling is twice that of reciprocal coupling, which means that nonreciprocal coupling is superior to reciprocal coupling in improving the measurement precision.

When the driving strength $\eta$ tends to be infinite,
the measurement uncertainty in the case of nonreciprocal case is given by

\begin{align}
{\delta g_{nr}}=\frac{2g^2\zeta[(\gamma_a+\kappa)^2+G_1^2)]}{G_1^2},
\label{eq:21}
\tag{21}
\end{align}
where $\zeta$ is a finite constant (see the detail in Appendix: C).

In the case of the reciprocal coupling with an infinite driving strength $\eta$, the expectation value of the cavity-field mode is given by
\begin{align}
|\alpha|=\sqrt{\frac{2G}{3\kappa}+\frac{\eta^{2/3}(\gamma_b^2+\omega_b^2)}{{(2\kappa^2\omega_b)}^{2/3} }}.
\label{eq:22}
\tag{22}
\end{align}
The expectation value of the homodyne detection with the operator $M$ is given by
\begin{align}
\langle M\rangle=\frac{2\eta}{2\kappa^2\omega_b/(\gamma_b^2+\omega_b^2])^{1/3}\eta^{2/3}-2\kappa\omega_bG/(\gamma_b^2+\omega_b^2)}.
\label{eq:23}
\tag{23}
\end{align}
The susceptibility is achieved in the case of the reciprocal coupling and the infinite driving intensity
\begin{align}
|\partial_g\langle M\rangle|_{\eta\rightarrow\infty}|=0.
\label{eq:24}
\tag{24}
\end{align}
It shows that the reciprocal coupling can not obtain any information of the gravitational acceleration in the case of infinite driving intensity $\eta$.
The intuitive explanation is that, in the case of reciprocal coupling, the infinitely large driving intensity makes the photon number of the cavity system tend to infinity. The force acting on the mechanical mode is much greater than gravity, making the gravitational acceleration information encoded on the cavity photonic system negligible. At this point, no information about the gravitational acceleration can be obtained by the homodyne detection.
In the case of non-reciprocal coupling, the mechanical mode will not be disturbed by the cavity photons. The information of the gravitational acceleration can be effectively transmitted to the cavity-field mode.

\textit{Two-photon driving}- We consider the cavity system subjected to an additional two-photon driving field, which is described by
\begin{align}
H_d=\frac{\chi}{2}(a^2e^{2i\omega_c t}+a^{2\dagger}e^{-2i\omega_c t}).
\label{eq:25}
\tag{25}
\end{align}
In the frame rotating at $\omega_c$, the corresponding quantum Langevin-Heisenberg equations of operators $a$ and $b$ are given by
\begin{align}
\dot{a}=&(-\gamma_a -\kappa)a-i\chi a^\dagger+2i\kappa a(b+b^\dagger)-i\eta+\sqrt{2\gamma_a} a_{in}\nonumber\\
&-i\sqrt{2\lambda}a(z_{in}+z^\dagger_{in}),\label{eq:26}
\tag{26}\\
\dot{b}=&(-i\omega_b-\gamma_b-\kappa )b,
+\sqrt{2\gamma_b} b_{in}+\sqrt{2\lambda}z_{in}-i G.
\label{eq:27}
\tag{27}
\end{align}

The critical point appears at $\chi_c=\pm\sqrt{(\gamma_a +\kappa)^2+G_1^2}$. When $-\chi_c<\chi<\chi_c$, the system can reach the steady state; otherwise, the system will not reach the steady state.
When $-\chi_c<\chi<\chi_c$, the mean value at the steady state is given by
\begin{align}
\alpha=&\frac{-i\eta[\kappa+\gamma_a-i(G_1-\chi)]}{(\gamma_a +\kappa)^2+G_1^2-\chi^2},\label{eq:28}
\tag{28}\\
\beta=&-i G/(i\omega_b+\gamma_b+\kappa )
\label{eq:29}.
\tag{29}
\end{align}

Near the critical point, the susceptibility of the gravitational acceleration $g$ is obtained (see the detail in Appendix: D)
\begin{align}
\partial_g\langle M\rangle|_{\chi\rightarrow\chi_c}=\frac{4\eta(\chi-G_1)G_1^2}{g[\chi_c^2-\chi^2]^2}.
\label{eq:30}
\tag{30}
\end{align}
It shows that the susceptibility tends to be infinite as the intensity of the driving field approaches the critical point.
The expectation value of the noise quantum  operator is given by
\begin{align}
{\langle\delta a^\dag\delta a\rangle_s}|_{\chi\rightarrow \chi_c}\propto(\chi_c^2-\chi^2)^{-3}.
\tag{31}
\end{align}
This means that the noise intensity will also tend to infinity at the critical point.

According to the error propagation formula, the measurement uncertainty of the gravitational acceleration near the critical point is described by
\begin{align}
\delta g_{\chi\rightarrow \chi_c}=\frac{\sqrt{2{\langle\delta a^\dag\delta a\rangle_s}|_{\chi\rightarrow \chi_c}+1}}{|\partial_g\langle M\rangle|_{\chi\rightarrow \chi_c}}\propto\sqrt{\chi_c^2-\chi^2}.
\tag{32}
\end{align}

In the case of reciprocal coupling, the quantum Langevin-Heisenberg equations of operators $a$ and $b$ are given by
\begin{align}
\dot{a}=&-\gamma_a a-i\chi a^\dagger+i\kappa a(b+b^\dagger) +\sqrt{2\gamma_a} a_{in}-i\eta,\label{eq:33}
\tag{33}\\
\dot{b}=&(-i\omega_b-\gamma_b)b+i\kappa a^\dagger a +\sqrt{2\gamma_b} b_{in}-i G.
\label{eq:34}
\tag{34}
\end{align}
The expectation value of the cavity-field mode at the steady is given by
\begin{align}
\alpha={\langle a\rangle}=\frac{-i\eta[\gamma_a-i(G_1-\kappa |\alpha|^2-\chi)]}{\gamma_a ^2+(G_1-\kappa |\alpha|^2)^2-\chi^2}.
\label{eq:35}
\tag{35}
\end{align}

The critical point appears at $\chi_{rc\pm}=\pm\sqrt{(\gamma_a )^2+(G_1-\kappa |\alpha|^2)^2}$. We define a quantity $\varepsilon$ to be equal to $\chi_{rc\pm}^2-\chi^2$, i.e., $\varepsilon=\chi_{rc\pm}^2-\chi^2$.
When $\chi$ is close to $\chi_{c\pm}$, the infinitesimal quantity $\varepsilon$ is close to 0. By calculating the eigenvalues of the evolution matrix, one can prove that the steady-state solution can exist only when $\chi$ approaches the negative critical point $\chi_{rc-}$. And there is no steady-state solution when $\chi$ approaches the positive critical point $\chi_{rc+}$.

In the case of the reciprocal coupling, the two-photon driving intensity $\chi$ is given by $\chi=\chi_{rc-}\approx\frac{-\kappa\eta^2\gamma_a^2}{\varepsilon^2}$.
This indicates that an infinitely large two-photon driving intensity is the prerequisite for achieving an infinitely large signal sensitivity. In contrast, in the case of non-reciprocal coupling, an extremely large signal sensitivity can be achieved without the need for an infinitely large driving intensity.

\textit{ The mechanical parametric amplification}-By a nonlinear modulation of the spring constant
of the mechanical resonator, the Hamiltonian of the mechanical parametric amplification is described by
\begin{align}
H_d=\frac{\upsilon}{2}b^2+\frac{\upsilon^*}{2}b^{2\dagger}.
\label{eq:36}
\tag{36}
\end{align}
When the  amplitude of the mechanical parametric amplification $\upsilon$ is close to the critical value $\upsilon_c=\pm\sqrt{(\gamma_a+\kappa)^2+\omega_b^2}$, the measurement susceptibility tends to zero in the case of the nonreciprocal coupling(see the detail in Appendix:E)
\begin{align}
|\partial_g\langle M\rangle|=0.
\label{eq:37}
\tag{37}
\end{align}
This indicates that the amplification of mechanical parameters does not improve the measurement precision of $g$, but rather reduces it.

In order to effectively utilize the amplification of mechanical parameters, an additional external force $f$ is required.
The Hamiltonian about the force $f$ is described by
\begin{align}
H_f=F(b+b^\dagger),
\label{eq:38}
\tag{38}
\end{align}
where $F=f\sqrt{\frac{1}{2m\omega_b}}$.
As a result, the combination of the amplification of mechanical parameters and the extra force can enhance the measurement susceptibility of the gravitational acceleration. Especially near the critical point, the measurement susceptibility tends to infinity. In the case of non-reciprocal coupling, the magnitude of the additional force does not require the knowledge of the photon number of the steady-state cavity field; while in the case of reciprocal coupling, the photon number of the steady-state cavity field needs to be known to obtain the appropriate magnitude of the additional force. Therefore, in the case of non-reciprocal coupling, the regulation of the additional force becomes easier.

Whether in the case of non-reciprocal or reciprocal coupling, the noise will also be amplified proportionally. Therefore, the measurement precision can not be improved by the combination of the amplification of mechanical parameters and the extra force.

\textit{Weak single-photon driving-}
In the previous section, we have explored that under a strong single-photon driving intensity, the mean-field approximation can be utilized. In this section, we investigate whether non-reciprocal coupling can further enhance the measurement precision of the gravitational acceleration under the weak single-photon driving limit: $\eta\ll1$.

Including the additional external force $f$, the optomechanical Hamiltonian is redescribed by
 \begin{align}
H=\omega a^\dagger a+\omega_m b^\dagger b-\kappa a^\dagger a(b^\dagger+ b)+G'(b^\dagger+ b),\label{eq:39}
\tag{39}
\end{align}
where $G'=G-F$. Our previous experience tells us that the smaller the value of $G'$ is, the better the measurement precision will be. Hence, we consider that $|G'|\ll1$.

Due to that the single-photon driving is weak and the gravitational force acting on the mechanical model is almost completely counterbalanced, the steady state of the entire system will be close to the ground state. Therefore, by ignoring the high-dimensional states, the optomechanical  state $|\phi(t)\rangle$ is described by the Fock state (see the detail in Appendix: F)
 \begin{align}
|\phi(t)\rangle=p_{00}|00\rangle+p_{01}|01\rangle+p_{10}|10\rangle+p_{11}|11\rangle,\label{eq:40}
\tag{40}
\end{align}
where $p_{ij}$ denote the probability amplitude of the Fock states $|ij\rangle$ with $i, j=0,1$.
The dissipative dynamical process can be dominated by the non-Hermitian Hamiltonian
 \begin{align}
H_w=&({\omega_b-i\gamma_b}) b^\dagger b-i\gamma_a a^\dagger a-\kappa a^\dagger a(b^\dagger+ b)\nonumber\\
&+G'(b^\dagger+ b)-i\lambda z^\dagger z+\eta(a+a^\dagger).\label{eq:41}
\tag{41}
\end{align}

When $\lambda=\kappa$, the non-Hermitian Hamiltonian is reduced to
 \begin{align}
H_{wn}=({\omega_m-i\gamma_b-i\kappa} )b^\dagger b-i\gamma_a a^\dagger a-2\kappa a^\dagger a b\nonumber\\
+G'(b^\dagger+ b)-i\kappa (a^\dagger a)^2+\eta(a+a^\dagger).\label{eq:42}
\tag{42}
\end{align}

The interaction Hamiltonian between the cavity-field mode and the mechanical mode becomes non-Hermitian
 \begin{align}
H_i=-2\kappa a^\dagger a b.\label{eq:43}
\tag{43}
\end{align}
It once again shows that the coupling between the cavity-field mode and the mechanical mode is nonreciprocal.

The optimal precision can be quantified by Cram\'{e}r-Rao bound\cite{lab36,lab37,lab38}
\begin{align}
\delta g\geq\sqrt{\frac{1}{\mathcal{F}}},\label{eq:44}
\tag{44}
\end{align}
where $\mathcal{F}$ denotes the quantum Fisher information. It can be calculated  by
\begin{align}
{\mathcal{F}}=4(|\partial_g|\phi\rangle_s|^2-|\langle\psi|\partial_g|\phi\rangle_s|^2),\label{eq:45}
\tag{45}
\end{align}
where the steady state $|\phi\rangle_s=|\phi(t\rightarrow\infty)\rangle$.

In the case of the nonreciprocal coupling, the measurement precision is
\begin{align}
\mathcal{F}_{nr}=\frac{8\kappa^2\eta^2{m}}{\omega_b(\kappa+\gamma_a)^4[(2\kappa+\gamma_a+\gamma_b)^2+\omega_b^2]}.\label{eq:46}
\tag{46}
\end{align}

In the reciprocal case, $\lambda=0$, the quantum Fisher information is given by
\begin{align}
\mathcal{F}_r=\frac{2\kappa^2\eta^2{m}}{\omega_b(\omega_b^2+\gamma_b^2)[(\kappa^2+\gamma_a^2+\gamma_a\gamma_b)^2+\gamma_a^2\omega_b^2]}.\label{eq:47}
\tag{47}
\end{align}

\begin{figure}[h]
\includegraphics[scale=0.35]{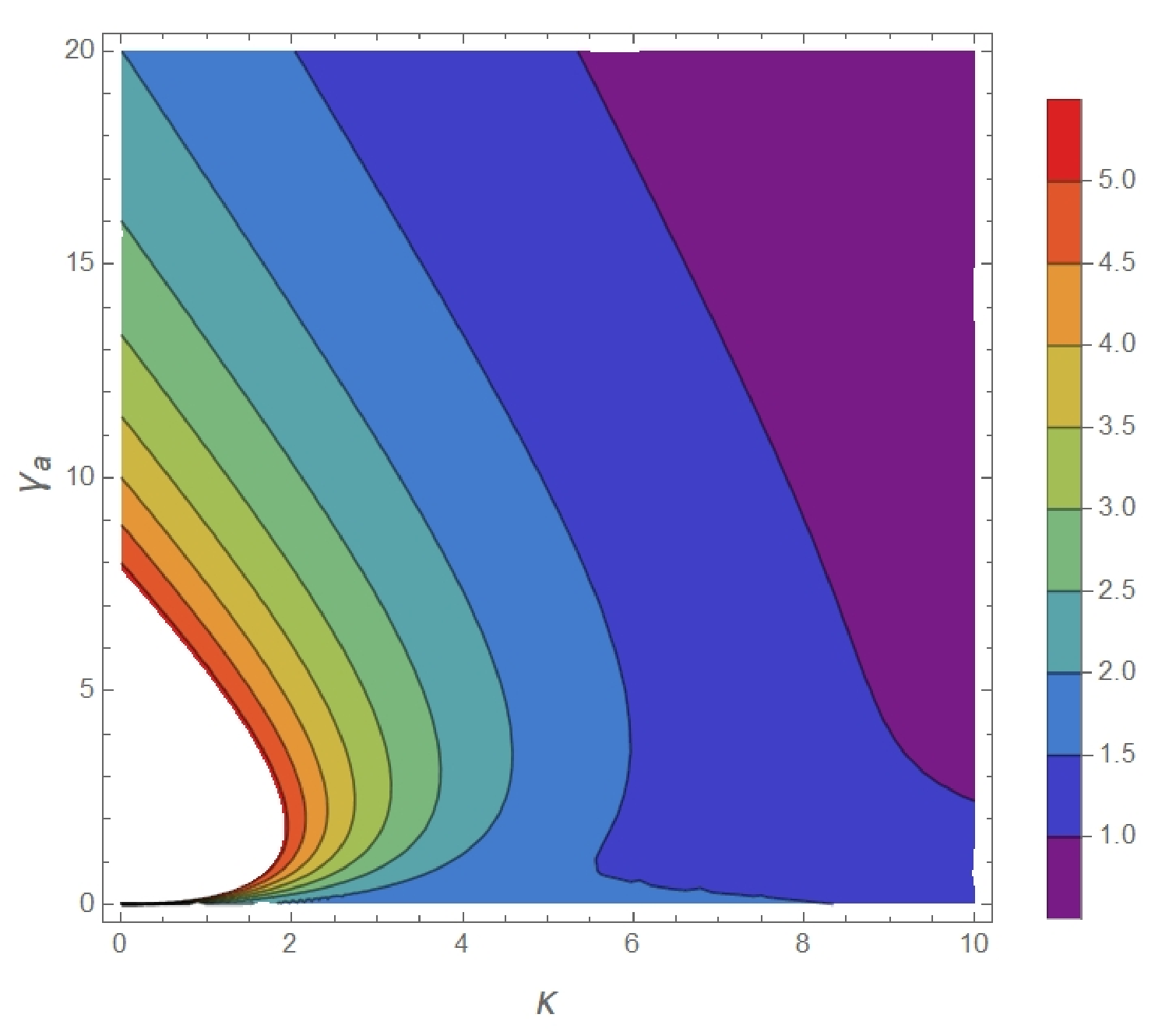}
 \caption{\label{fig.2} Graph showing the ratio $R_w$ of the optimal measurement uncertainty obtained by nonreciprocal coupling and reciprocal coupling varies with the coupling strength $\kappa$ and the cavity field dissipation rate $\gamma_a$. Different colors represent different values of the ratio $R_w$. Here, the dimensionless parameters are given by: $\gamma_b=1$, and $\omega_b=20$.}
\end{figure}

When $\kappa\ll\gamma_a$ and $\kappa\ll\gamma_b$, the ratio of the measurement precision obtained through non-reciprocity to that through reciprocity is
\begin{align}
R_w=\frac{2\sqrt{\omega_b^2+\gamma_b^2}}{\gamma_a}\approx\frac{2\omega_b}{\gamma_a},\label{eq:48}
\tag{48}
\end{align}
where the approximation in the latter part stems from the fact that in most cases, the dissipation rate of the mechanical model is much lower than the frequency of the mechanical model.
The above equation can show that the nonreciprocal coupling can perform more than twice as well as the reciprocal coupling when the dissipation rate is less than the frequency of the mechanical resonator. As shown in Fig.~\ref{fig.2}, the ratio $R_w$ is greater than 1 in most parameter cases. This indicates that within a wide range of experimental parameters, the performance of non-reciprocal coupling is superior to that of reciprocal coupling.

\textit{{Conclusion and outlook}}-We construct the nonreciprocal optomechanical coupling to sense the gravitational acceleration $g$. The results show that in many cases, non-reciprocal coupling performs better than reciprocal coupling in improving the measurement precision of gravitational parameters. The nonreciprocal coupling can achieve a two-fold improvement in the measurement precision in the case of the strong  single-photon driving and weak coherent coupling. As the single-photon driving intensity approaches infinity, the measurement uncertainty in the non-reciprocal coupling case stabilizes at a finite value, while in the reciprocal coupling case, the measurement uncertainty stabilizes at an infinite value. In the case of nonreciprocal coupling, the measurement uncertainty tends to zero as the two-photon driving strength approaches the finite critical value. However, in order to achieve a similar result for the reciprocal coupling situation, an infinitely large two-photon driving intensity is required. In the case of the weak single-photon driving, non-reciprocal coupling still
performs better than reciprocal coupling under most experimental parameters. Our scheme is feasible in different experimental platforms\cite{lab22,lab23,lab24,lab24,lab25,lab26,lab26,lab27}. And by synthetic magnetism and reservoir
 engineering, a generalized non-reciprocity in an optomechanical circuit has been observed in the experiment\cite{lab39}. Our work will open up the way to enhance the measurement precision of the gravitational acceleration by making use of nonreciprocal coupling. It is meaningful to extend the nonreciprocal coupling from the optomechanical system to the spin-mechanics system\cite{lab12}.
\section*{Acknowledgements}
This research was supported by the National Natural Science Foundation of China (Grant No. 12365001 and No. 62001134), the Bagui Youth Top Talent Training Program, and Guangxi Natural Science Foundation (Grant No. 2020GXNSFAA159047).

\section*{Appendix A: Dissipative coupling}
The interaction Hamiltonian between the optomecahnical system and the mode $c$ is described as
\begin{align}
H_{I}=\mu a^\dagger a(c+c^\dagger)+\nu b^\dagger c+\nu^* b c^\dagger.
\label{eq:A1}
\tag{A1}
\end{align}
The master equation of the modes $a,b,c$ is described by
\begin{align}
\dot{\rho}_{abc}=-i[H+H_{I}+H_c,\rho]+\sum_{j=a,b,c}\gamma_j\mathcal{D}_j[\rho_{abc}],\tag{A2}\label{eq:A2}
\end{align}
where $H_c=\omega_c c^\dagger c$ denotes the Hamiltonian of the mode $c$ with  the frequency $\omega_c$.
The model $c$ suffers from a large local dissipation $\gamma_c$.
The quantum Langevin-Heisenberg equation of the operator $c$ is given by
\begin{align}
\dot{c}=(-i\omega_c-\gamma_c )c-i\mu a^\dagger a-i \nu b +\sqrt{2\gamma_c} c_{in}.
\label{eq:A3}
\tag{A3}
\end{align}
Assuming that $\gamma_c\gg \gamma_a, \gamma_b$, the mode $c$ can be adiabatically eliminated. Let $\dot{c}=0$, we get that
\begin{align}
{c}=(-i\mu a^\dagger a-i \nu b +\sqrt{2\gamma_c} c_{in})/(i\omega_c+\gamma_c ).
\label{eq:A4}
\tag{A4}
\end{align}
Let $\nu=\omega_c-i\gamma_c$ and substitute the above equation into the master equation Eq.~(\ref{eq:A2}), we can derive the reduced master equation
\begin{align}
\dot{\rho}=-i[H,\rho]+\sum_{j=a,b}\kappa_j\mathcal{D}_j[\rho]+\frac{\mu^2}{\omega_c^2+\gamma_c^2}\mathcal{D}_z[\rho].\tag{A5}
\end{align}
Set $\frac{\mu^2}{\omega_c^2+\gamma_c^2}=\lambda$, the dissipative coupling in Eq.~(\ref{eq:2}) of the main text can be obtained.
\section*{Appendix B: Nonreciprocal coupling}
The evolution equations of quantum fluctuations are given by
\begin{align}
\dot{\delta a}=&[-\gamma_a -\kappa+2i\kappa (\beta+\beta^*)]\delta a-2i\kappa \alpha( \delta b+\delta b^\dagger)\nonumber\\
 &+\sqrt{2\gamma_a} a_{in}-i\sqrt{2\lambda}\alpha(z_{in}+z^\dagger_{in},)\tag{B1}\\
\dot{\delta b}=&(-i\omega_b-\gamma_b-\kappa )\delta b +\sqrt{2\gamma_b} b_{in}+\sqrt{2\kappa}z_{in}.\tag{B2}
\end{align}

The above equations can be restated as
\begin{align}
\mathbf{\dot{A}}= \mathbb{M}\mathbf{A}+\mathbf{A}_{in}
\label{eq:A8}
\tag{B3}
\end{align}
in which,
\[
 \mathbb{M}=\left(
\begin{array}{ll}
 -\gamma_a-\kappa-i\theta\ \ \ \ \ \ 0\ \ \ \ \ -2i\kappa \alpha\ \ \ -2i\kappa \alpha\\
 \ \ \ \ \ \ 0\ \ \ \ \ -\gamma_a-\kappa+i\theta\ \ \ 2i\kappa \alpha^*\ \ \ \ \ 2i\kappa \alpha^*\\
 \ \ \  \ \ \ 0\ \ \ \ \ \ \ \ \ \ \ \ \ \ \ \ 0\ \ \ -i\omega_b-\gamma_b-\kappa\ \ \ \ 0\\
 \ \ \ \ \ \ 0\ \ \ \ \ \ \ \ \ \ \ \ \ \ \ \ 0\ \ \  \ \ \ \ \ \ \ 0\ \ \ \ \ i\omega_b-\gamma_b-\kappa
  \end{array}
\right ).\tag{B4}\]
\label{eq:S9}
where $\theta=-2\kappa(\beta+\beta^*)$, $\mathbf{A}=\{\delta a,\delta a^\dagger,\delta b, \delta b^\dagger\}$
$\mathbf{A}_{in}=\{A_{1in},A^\dagger_{1in},A_{2in},A^\dagger_{2in}\}$, where the noise operators $A_{1in}=\sqrt{2\gamma_a}a_{in}-i\sqrt{2\kappa}\alpha(z_{in}+z^\dagger_{in})$, $A_{2in}=\sqrt{2\gamma_b}b_{in}+\sqrt{2\kappa}z_{in}.$
The solution is given by
\begin{align}
\mathbf{{A}}(t)= \exp[\mathbb{M}t]\mathbf{A}(0)+\int_0^tds\exp[\mathbb{M}(t-s)]\mathbf{A}_{in}(s).
\label{eq:3}
\tag{B5}
\end{align}
According to the above solution, we can obtain
\begin{align}
{\delta a}(t)=&\exp\{[-\gamma_a -\kappa+2i\kappa (\beta+\beta^*)]t\}\delta a(0)+\nonumber\\
&\int_0^tds\exp\{[-\gamma_a -\kappa+2i\kappa (\beta+\beta^*)](t-s)\}\nonumber\\
&\times\{-2i\kappa \alpha[ \delta b(s)+\delta b^\dagger(s)]+\sqrt{2\gamma_a} a_{in}(s)\nonumber\\
&-i\sqrt{2\kappa}\alpha(z_{in}(s)+z^\dagger_{in}(s))\},\tag{B6}\\
\delta b(t)=&\exp[(-i\omega_b-\gamma_b-\lambda )t]\delta b(0) +\nonumber\\
&\int_0^tds\exp[(-i\omega_b-\gamma_b-\kappa )(t-s)]\nonumber\\
&\times[\sqrt{2\gamma_b} b_{in}(s)+\sqrt{2\lambda}z_{in}(s)].
\tag{B7}
\end{align}
The cavity-field mode at the steady state is described by
\begin{align}
&\delta a(t\rightarrow\infty)=\int_0^\infty dt\{\exp[-t(\kappa+\gamma_a+i\theta)]A_{1in}(t)+\nonumber\\
&\frac{2\alpha \kappa \{\exp[(-\kappa-\gamma_b+i\omega_b)t]-\exp[(-\kappa-\gamma_a-i\theta)t]\} A_{2in}(t)}{i(\gamma_a-\gamma_b)+\omega_b-\theta}+\nonumber\\
&\frac{2\alpha \kappa\{\exp[(-\kappa-\gamma_a-i\theta)t]-\exp[(-\kappa-\gamma_b-i\omega_b)t]\}A^\dagger_{2in}(t)}{i(-\gamma_a+\gamma_b)+\omega_b+\theta}\}.
\label{eq:3}
\tag{B8}
\end{align}

The expectation value of the cavity-field mode at the steady state is described by
\begin{widetext}
\begin{align}
\langle\delta a^\dag(t\rightarrow\infty)\delta a(t\rightarrow\infty)\rangle=
\frac{\kappa|\alpha|^2[(\kappa+\gamma_b)[(2\kappa+\gamma_a+\gamma_b)^2+(\omega_b+\theta)^2]+4\kappa(\kappa+\gamma_b)(2\kappa+\gamma_a+\gamma_b)+8\kappa^2(2\kappa+\gamma_a+\gamma_b)]}
{(\kappa+\gamma_a)(\kappa+\gamma_b)(2\kappa+\gamma_a+\gamma_b)^2+(\omega_b+\theta)^2}.
\label{eq:3}
\tag{B9}
\end{align}
\end{widetext}

\section*{Appendix C: Reciprocal coupling}
When there is no dissipative coupling, i.e., $\lambda=0$, the quantum Langevin-Heisenberg equations are given by
\begin{align}
\dot{a}&=-\gamma_a a+i\kappa a(b+b^\dagger) +\sqrt{2\gamma_a} a_{in}-i\eta,\tag{C1}\\
\dot{b}&=(-i\omega_b-\gamma_b)b+i\kappa a^\dagger a +\sqrt{2\gamma_b} b_{in}-i G.
\label{eq:3}
\tag{C2}
\end{align}
Using the mean-field approximation, the evolution equations of the expectation values are
\begin{align}
\dot{\langle a\rangle}&=(-\gamma_a )\langle a\rangle+i\kappa \langle a \rangle\langle b+b^\dagger\rangle-i\eta,\tag{C3}\\
\dot{\langle b\rangle}&=(-i\omega_b-\gamma_b )\langle b\rangle -i G+i \kappa|\langle a\rangle|^2.
\label{eq:3}
\tag{C4}
\end{align}

Let the left terms be equal to 0, we can obtain the expectation values at the steady state
\begin{align}
\langle a\rangle&=\alpha=\frac{i \eta}{-\gamma_a+\frac{2i\kappa \omega_b(\kappa|\alpha|^2-G )}{\gamma_b^2+\omega_b^2}},\tag{C5}\\
\langle b\rangle&=\beta=\frac{i|\alpha|^2-iG}{i\omega_b+\gamma_b},\tag{C6}\\
|\alpha|^2&=\frac{\eta^2}{\gamma_a^2+\frac{4\kappa^2 \omega_b^2(\kappa|\alpha|^2-G )^2}{(\gamma_b^2+\omega_b^2)^2}}.
\label{eq:C7}
\tag{C7}
\end{align}

When $\kappa |\alpha|^2\ll G$, we can obtain that
\begin{align}
\alpha&=\frac{i \eta}{-\gamma_a+\frac{2i\kappa \omega_b(-G )}{\gamma_b^2+\omega_b^2}},\tag{C8}\\
|\alpha|^2&=\frac{\eta^2}{\gamma_a^2+\frac{4\kappa^2 \omega_b^2(G )^2}{(\gamma_b^2+\omega_b^2)^2}}.
\label{eq:3}
\tag{C9}
\end{align}

When $\kappa |\alpha|^2\ll G$, we can obtain the measurement uncertainty of the gravitational acceleration in the case of reciprocal and non-reciprocal cases
\begin{align}
\delta g_{r}&=\frac{2g(\gamma_a ^2+G_1^2)}{\eta G_1},\tag{C10}\\
\delta g_{nr}&=\frac{g(\gamma_a^2+G_1^2)}{\eta G_1}.
\label{eq:3}
\tag{C11}
\end{align}

The ratio of the uncertainties of the gravitational acceleration measured in reciprocal and nonreciprocal cases is
\begin{align}
R=\frac{\delta g_{nr}}{\delta g_r}=1/2.
\label{eq:3}
\tag{C12}
\end{align}

When the driving strength $\eta$ tends to be infinite,
the measurement uncertainty in the case of nonreciprocal case is given by

\begin{align}
{\delta g_{nr}}=\frac{2g^2\zeta[(\gamma_a+\kappa)^2+G_1^2]}{G_1^2},
\label{eq:3}
\tag{C13}
\end{align}
in which,
\begin{widetext}
\begin{align}
\zeta&=
\frac{\kappa[(\kappa+\gamma_b)[(2\kappa+\gamma_a+\gamma_b)^2+(\omega_b+\theta)^2]+4\kappa(\kappa+\gamma_b)(2\kappa+\gamma_a+\gamma_b)+8\kappa^2(2\kappa+\gamma_a+\gamma_b)]}
{(\kappa+\gamma_a)(\kappa+\gamma_b)[(2\kappa+\gamma_a+\gamma_b)^2+(\omega_b+\theta)^2]},\tag{C14}\\
&G_1=4G\kappa \omega_b/[(\gamma_b+\kappa)^2+\omega_b^2].
\label{eq:3}
\tag{C15}
\end{align}
\end{widetext}

In the case of the reciprocal coupling and infinite $\eta$, the solution of Eq.~(\ref{eq:C7}) is given by
\begin{align}
|\alpha|^2=\frac{2G}{3\kappa}+\frac{\eta^{2/3}(\gamma_b^2+\omega_b^2)}{{(2\kappa^2\omega_b)}^{2/3} }.
\tag{C16}
\end{align}

The expectation value of the homodyne detection is given by
\begin{align}
\langle M\rangle=\frac{2\eta}{[(2\kappa^2\omega_b/(\gamma_b^2+\omega_b^2)]^{1/3}\eta^{2/3}-2\kappa\omega_bG/(\gamma_b^2+\omega_b^2)}.
\tag{C17}
\end{align}
Then, by differentiating the above equation, we can obtain
\begin{align}
\partial_g\langle M\rangle|_{\eta\rightarrow\infty}=0.
\tag{C18}
\end{align}

\section*{Appendix D: Two-photon driving}
Including the two-photon driving, the evolution matrix can be described by
\begin{align}
\mathbf{\dot{A}}= \mathbb{M}\mathbf{A}+\mathbf{A}_{in},
\label{eq:D1}
\tag{D1}
\end{align}
in which,
\[
 \mathbb{M}=\left(
\begin{array}{ll}
 -\gamma_a-\kappa-iG_1 \ \ \ \ \ -i\chi\ \ \ \ \ \ -2i\kappa \alpha\ \ \ \ -2i\kappa \alpha\\
\ \ \ \ \ \ i\chi\ \ \ \ \ \ \ \ -\gamma_a-\kappa+iG_1\ \ 2i\kappa \alpha^*\ \ \ \ \ \ 2i\kappa \alpha^*\\
\ \ \ \ \ \ 0\ \ \ \ \ \ \ \ \ \ \ \ \ \ \ \ \ \ \ \ \ 0\ \ \ -i\omega_b-\gamma_b-\kappa\ \ \ 0\\
\ \ \ \ \ \ 0\ \ \ \ \ \ \ \ \ \ \ \ \ \ \ \ \ \ \ \ \ 0\ \ \ \ \ \ \ \ \ 0\ \ \ \ i\omega_b-\gamma_b-\kappa
  \end{array}
\right ).\tag{D2}\]
\label{eq:D2}The noise operators are
$\mathbf{A}_{in}=\{A_{1in},A^\dagger_{1in},A_{2in},A^\dagger_{2in}\}$, where $A_{1in}=\sqrt{2\gamma_a}a_{in}-i\sqrt{2\kappa}\alpha(z_{in}+z^\dagger_{in})$, and $A_{2in}=\sqrt{2\gamma_b}b_{in}+\sqrt{2\kappa}z_{in}.
$
The eigenvalues of the evolution matrix are given by :$\pm i\omega_b-\gamma_b-\kappa, -\gamma_a-\kappa\pm\sqrt{\chi^2-G_1^2}$.
The condition for the system to reach a steady state is that all eigenvalues of the evolution matrix should be negative.
The critical point appears at $\chi_c=\pm\sqrt{(\gamma_a +\kappa)^2+G_1^2}$. When $-\chi_c<\chi<\chi_c$, the system can reach the steady state; otherwise, the system will not reach a steady state.

When the two-photon driving strength $\chi$ is close to the critical values $\chi_c$, we can obtain
\begin{align}
{\langle\delta a^\dag\delta a\rangle_s}|_{\chi\rightarrow \chi_c}\propto[(\gamma_a +\kappa)^2+G_1^2-\chi^2]^{-3}.
\tag{D3}
\end{align}

The measurement uncertainty of the gravitational acceleration is described by
\begin{align}
\delta g_{\chi\rightarrow \chi_c}&=\frac{\sqrt{2{\langle\delta a^\dag\delta a\rangle_s}|_{\chi\rightarrow \chi_c}+1}}{|\partial_g\langle M\rangle|_{\chi\rightarrow \chi_c}}\nonumber\\
&\propto[(\gamma_a +\kappa)^2+G_1^2-\chi^2]^{1/2}.
\tag{D4}
\end{align}

\section*{Appendix E: Mechanical parametric amplification}
In the frame rotating at $\omega_c=\omega_d$, including the mechanical parametric amplification, the corresponding quantum Langevin-Heisenberg equations of operators $a$ and $b$ are given by
\begin{align}
\dot{a}=&(-\gamma_a -\kappa)a+2i\kappa a(b+b^\dagger) \nonumber\\
&+\sqrt{2\gamma_a} a_{in}-i\eta-i\sqrt{2\lambda}a(z_{in}+z^\dagger_{in}),\tag{E1}\\
\dot{b}=&(-i\omega_b-\gamma_b-\kappa )b-i\upsilon b^\dagger\nonumber\\
&+\sqrt{2\gamma_b} b_{in}+\sqrt{2\lambda}z_{in}-i G.
\label{eq:3}
\tag{E2}
\end{align}

The mean value at the steady state is given by
\begin{align}
\alpha&={\langle a\rangle}=\frac{-i\eta}{\gamma_a+\kappa-iG_2},\tag{E3}\\
\beta&={\langle b\rangle}=\frac{G[\lambda-\omega_b-i(\kappa+\gamma_b)]}{(\kappa+\gamma_b)^2+\omega_b^2-\upsilon^2},
\label{eq:3}
\tag{E4}
\end{align}
where the phase factor $G_2=\frac{4G\kappa(\upsilon-\omega_b)}{(\gamma_a+\kappa)^2+\omega_b^2-\upsilon^2}$.
Then, by differentiating the above equation, we can obtain
\begin{align}
|\partial_g\langle M\rangle|=\frac{2\eta G_2}{g[(\gamma_a+\kappa)^2+G_2^2]}-\frac{4\eta G_2^3}{g[(\gamma_a+\kappa)^2+G_2^2]^2}
\label{eq:3}
\tag{E5}
\end{align}

When the driving strength $\upsilon$ is close to the critical value $\upsilon_c=\pm\sqrt{(\gamma_a+\kappa)^2+\omega_b^2}$, the phase factor tends to be infinite. This further leads to the measurement susceptibility approaching zero, i.e.,
\begin{align}
|\partial_g\langle M\rangle|=0.
\label{eq:3}
\tag{E6}
\end{align}
This indicates that the amplification of mechanical parameters not only fails to enhance the measurement precision of $g$, but actually reduces it.

In order to effectively utilize the amplification of mechanical parameters, an additional external force $f$ is required.
The Hamiltonian about the force $f$ is described by
\begin{align}
H_f=F(b+b^\dagger).
\label{eq:3}
\tag{E7}
\end{align}
where $F=f\sqrt{\frac{1}{2m\omega_b}}$.

Including the Hamiltonian $H_f$, the quantum Langevin-Heisenberg equations of operators $a$ and $b$ are rewritten as
\begin{align}
\dot{a}=&(-\gamma_a -\kappa)a+2i\kappa a(b+b^\dagger) \nonumber\\
&+\sqrt{2\gamma_a} a_{in}-i\eta-i\sqrt{2\lambda}a(z_{in}+z^\dagger_{in})\tag{E8}\\
\dot{b}=&(-i\omega_b-\gamma_b-\kappa )b-i\upsilon b^\dagger\nonumber\\
&+\sqrt{2\gamma_b} b_{in}+\sqrt{2\lambda}z_{in}-i (G-F).
\tag{E9}
\end{align}
The mean values at the steady state are given by
\begin{align}
\alpha=&{\langle a\rangle}=\frac{-i\eta}{\gamma_a+\kappa-iG_3},\tag{E10}\\
\beta=&{\langle b\rangle}=\frac{(G-F)[\lambda-\omega_b-i(\kappa+\gamma_b)]}{(\kappa+\gamma_b)^2+\omega_b^2-\upsilon^2},
\tag{E11}
\end{align}
where the phase factor $G_3=\frac{4(G-F)\kappa(\upsilon-\omega_b)}{(\gamma_a+\kappa)^2+\omega_b^2-\upsilon^2}$.
The susceptibility is derived
\begin{align}
|\partial_g\langle M\rangle|=\frac{2\eta G_2}{g[(\gamma_a+\kappa)^2+G_3^2]}-\frac{4\eta G_2G_3^2}{g[(\gamma_a+\kappa)^2+G_3^2]^2}.
\tag{E12}
\end{align}

When $G=F$, the value $G_3=0$, one can obtain
\begin{align}
|\partial_g\langle M\rangle|=\frac{2\eta G_2}{g(\gamma_a+\kappa)^2}.
\tag{E13}
\end{align}
At this point, when the driving intensity of mechanical parameter approaches the critical value, the measurement sensitivity is close to be infinite.

The reciprocal coupling:
In the case of reciprocal coupling, the quantum Langevin-Heisenberg equations of operators $a$ and $b$ are given by
\begin{align}
\dot{a}&=-\gamma_a a+i\kappa a(b+b^\dagger) +\sqrt{2\gamma_a} a_{in}-i\eta,\tag{E14}\\
\dot{b}&=(-i\omega_b-\gamma_b)b-i\upsilon b^\dagger+i\kappa a^\dagger a +\sqrt{2\gamma_b} b_{in}-i (G-F).
\tag{E15}
\end{align}
The mean value at the steady state is given by
\begin{align}
\alpha&={\langle a\rangle}=\frac{-i\eta}{\gamma_a-iG_4},\tag{E16}\\
\beta&={\langle b\rangle}=\frac{(G-F-\kappa |\alpha|^2 )(\upsilon-\omega_b-i\gamma_b)}{\gamma_b^2+\omega_b^2-\upsilon^2},
\tag{E17}
\end{align}
where the phase factor $G_4=\frac{2(G-F-\kappa |\alpha|^2)\kappa(\upsilon-\omega_b)}{(\gamma_a)^2+\omega_b^2-\upsilon^2}$.
By squaring the absolute value of $\alpha$, we obtain
\begin{align}
|\alpha|^2=\frac{\eta^2}{\gamma_a^2+G_4^2}.
\tag{E18}
\end{align}
When $G=F$, the above equation has no real solutions, which means there is no steady state.
When $F=G-\kappa|\alpha|^2$ and $\upsilon\rightarrow\upsilon_c$, we can obtain the solution
\begin{align}
|\alpha|^2=\frac{\eta^2}{\gamma_a^2}
\tag{E19}
\end{align}
The measurement susceptibility is given by
\begin{align}
|\partial_g\langle M\rangle|=\frac{\eta G_2}{g\gamma_a^2}.
\tag{E20}
\end{align}

By following the similar steps mentioned earlier, we can obtain the expectation value of the noise operator
\begin{align}
{\langle\delta a^\dag\delta a\rangle_s}|_{\upsilon\rightarrow \upsilon_c}\propto 1/[(\gamma_a+\kappa)^2+\omega_b^2-\upsilon^2]^2.
\tag{E21}
\end{align}
The measurement uncertainty is described by
\begin{align}
\delta g_{\chi\rightarrow \upsilon_c}=\frac{\sqrt{2{\langle\delta a^\dag\delta a\rangle_s}|_{\chi\rightarrow \upsilon_c}+1}}{|\partial_g\langle M\rangle|_{\chi\rightarrow \upsilon_c}}=C,
\tag{E22}
\end{align}
where the value of $C$ is independent of the $(\gamma_a+\kappa)^2+\omega_b^2-\upsilon^2$. It shows that the combination of the amplification of mechanical parameters and the extra force can not improve the measurement precision of $g$,  but only improve the measurement susceptibility.

\section*{Appendix F: Weak single-photon driving}
According to the Schr\"{o}dinger equation, $i|\dot{\psi}(t)\rangle=H_{w}|{\psi(t)}\rangle$, we can obtain the evolution equations of the probability amplitudes:
 \begin{align}
\dot{p_{00}}&=p_{01}G'+p_{10}\eta,\tag{F1}\\
\dot{p_{01}}&=p_{01}(\omega_b-i\gamma_b)+p_{11}\eta,\tag{F2}\\
\dot{p_{10}}&=p_{00}\eta-ip_{10}(\kappa+\gamma_a)+p_{11}(G'-2\kappa),\tag{F3}\\
\dot{p_{11}}&=p_{01}\eta+p_{10}G'+p_{11}(\omega_b-i\gamma_b-2i\kappa-i\gamma_a).\tag{F4}
\end{align}

Let the probability amplitude $p_{00}$ at the steady state $p^s_{00}=1$ and $\dot{p_{01}}=\dot{p_{10}}=\dot{p_{11}}=0$, one can obtain the values of the probability amplitudes at the steady state

 \begin{align}
{p^s_{01}}&=0+O(\eta^2),\tag{F5}\\
{p^s_{10}}&=\frac{ \eta(\omega_b- i \Gamma_b) (2 \kappa +i\omega_b +\gamma_a +\gamma_b)}{  (i \omega_b + \gamma_b) [ -
    2 G' \kappa + (\kappa + \gamma_a) (2 \kappa +
       i \omega_b + \gamma_a + \gamma_b)]},\tag{F6}\\
{p^s_{11}}&=0+O(\eta G').\tag{F7}
\end{align}

The steady state of the cavity field mode can be described by
\begin{align}
|\phi\rangle_s=|0\rangle+{p^s_{10}}|1\rangle.\tag{F8}
\end{align}

In the reciprocal case, $\lambda=0$, the corresponding evolution equations of the probability amplitudes are
 \begin{align}
\dot{p_{00}}&=p_{01}G'+p_{10}\eta,\tag{F9}\\
\dot{p_{01}}&=p_{01}(\omega_b-i\gamma_b)+p_{11}\eta+p_{00}G,\tag{F10}\\
\dot{p_{10}}&=p_{00}\eta-ip_{10}\gamma_a+p_{11}(G'-\kappa),\tag{F11}\\
\dot{p_{11}}&=p_{01}\eta+p_{10}(G'-\kappa)+p_{11}(\omega_b-i\gamma_b-i\gamma_a).\tag{F12}
\end{align}
Let $p^s_{00}=1$ and $\dot{p_{01}}=\dot{p_{10}}=\dot{p_{11}}=0$, one can obtain the values at the steady state

 \begin{align}
{p^s_{01}}&=\frac{-G'}{\omega_b-i\gamma_b}+O(\eta^2)\tag{F13}\\
{p^s_{10}}&=\frac{\eta\{G'\kappa-(\omega_b-i\gamma_b)[\omega_b-i(\gamma_a+\gamma_b)]\}}{(\omega_b-i\gamma_b)\{\kappa^2+\gamma_a[i\omega_b+(\gamma_a+\gamma_b)]\}}\tag{F14}\\
{p^s_{11}}&=\frac{\kappa\eta}{\kappa^2+\gamma_a[i\omega_b+(\gamma_a+\gamma_b)]}+O(\eta G').\tag{F15}
\end{align}

\end{document}